\DocumentMetadata{}
\documentclass[sigconf]{acmart}
\makeatletter
\def\@ACM@checkaffil{
    \if@ACM@instpresent\else
    \ClassWarningNoLine{\@classname}{No institution present for an affiliation}%
    \fi
    \if@ACM@citypresent\else
    \ClassWarningNoLine{\@classname}{No city present for an affiliation}%
    \fi
    \if@ACM@countrypresent\else
        \ClassWarningNoLine{\@classname}{No country present for an affiliation}%
    \fi
}
\makeatother

\AtBeginDocument{%
  }

\copyrightyear{2026}
\acmYear{2026}
\setcopyright{cc}
\setcctype{by}
\acmConference[WWW '26] {Proceedings of the ACM Web Conference 2026}{April 13--17, 2026}{Dubai, United Arab Emirates.}
\acmBooktitle{Proceedings of the ACM Web Conference 2026 (WWW '26), April 13--17, 2026, Dubai, United Arab Emirates}
\acmISBN{979-8-4007-2307-0/2026/04}
\acmDOI{10.1145/3774904.3792872}

\usepackage{booktabs}
\usepackage{multirow}
\usepackage{url}
\usepackage{enumitem}   
\usepackage{float} 
\usepackage{hyperxmp}

\definecolor{myred}{rgb}{0.68627451, 0.14117647, 0.09803922}

\newcommand{\myeq}[1]{\hyperref[eq:#1]{Eq.~(\ref*{eq:#1})}}
\newcommand{\mysec}[1]{\hyperref[sec:#1]{Section~\ref*{sec:#1}}}
\newcommand{\mytable}[1]{\hyperref[tab:#1]{Table~\ref*{tab:#1}}}
\newcommand{\myfig}[1]{\hyperref[fig:#1]{Fig.~\ref*{fig:#1}}}
\newcommand{\myappendix}[1]{\hyperref[appendix:#1]{Appendix}}
\newcommand{\myalg}[1]{\hyperref[alg:#1]{Algorithm~\ref*{alg:#1}}}
\newcommand{\myhyp}[1]{\hyperref[alg:#1]{\textsc{Hypothesis}~\ref*{hyp:#1}}}

\usepackage{array}

\begin{document}

\title{LLM Reasoning for Cold-Start Item Recommendation}



\author{Shijun Li}
\authornote{Work was done while Shijun was interning at Netflix.}
\email{shijunli@utexas.edu}
\affiliation{%
  \institution{The University of Texas at Austin}
  \city{Austin}
  \country{United States}
}

\author{Yu Wang}
\authornote{Work was done while Yu was working at Netflix.}
\email{yu.wang1@capitalone.com}
\affiliation{%
  \institution{Capital One AI Foundations}
  \city{Los Gatos}
  \country{United States}
}

\author{Jin Wang}
\email{jinw@netflix.com}
\affiliation{%
  \institution{Netflix}
  \city{Los Gatos}
  \country{United States}
}

\author{Ying Li}
\email{yingl@netflix.com}
\affiliation{%
  \institution{Netflix}
  \city{Los Gatos}
  \country{United States}
}

\author{Joydeep Ghosh}
\email{jghosh@utexas.edu}
\affiliation{%
  \institution{The University of Texas at Austin}
  \city{Austin}
  \country{United States}
}

\author{Anne Cocos}
\email{acocos@netflix.com}
\affiliation{%
  \institution{Netflix}
  \city{Los Gatos}
  \country{United States}
}


\begin{abstract}

Large Language Models (LLMs) have shown significant potential for improving recommendation systems through their inherent reasoning capabilities and extensive knowledge base. Yet, existing studies predominantly address warm-start scenarios with abundant user-item interaction data, leaving the more challenging cold-start scenarios, where sparse interactions hinder traditional collaborative filtering methods, underexplored. To address this limitation, we propose novel reasoning strategies designed for cold-start item recommendations within the Netflix domain. Our method utilizes the advanced reasoning capabilities of LLMs to effectively infer user preferences, particularly for newly introduced or rarely interacted items. We systematically evaluate supervised fine-tuning, reinforcement learning-based fine-tuning, and hybrid approaches that combine both methods to optimize recommendation performance. Extensive experiments on real-world data demonstrate significant improvements in both methodological efficacy and practical performance in cold-start recommendation contexts. Remarkably, our reasoning-based fine-tuned models outperform Netflix's production ranking model by up to 8\% in certain cases.


\end{abstract}

\maketitle

\section{Introduction}

The application of Large Language Models (LLMs) to recommendation systems has emerged as a promising research direction, drawing significant interest from both academia and industry~\cite{carraro2024enhancing, liu2024rec}. Recent studies highlight the potential of LLM reasoning to further enhance recommendation quality~\cite{wang2023drdt,liu2025improving}. However, these approaches have mainly been evaluated in warm-start scenarios with sufficient historical interactions available for the recommended items.

In practical settings, cold-start scenarios pose a significant challenge for recommender systems \cite{li2021seamlessly, yang2025cold}, as classic collaborative filtering methods typically underperform due to limited interactions. In contrast, LLMs which equipped from extensive world knowledge and advanced reasoning abilities, are particularly well-positioned to address these challenges by inferring user preferences for new or infrequently interacted items. Despite this potential, current research on LLM for recommendation largely focuses on direct reasoning or reasoning adjusted by supervised fine-tuning, leaving reinforcement learning-based strategies, which are typically considered as more promising and flexible, relatively underexplored.

To address these gaps, we develop and validate novel reasoning strategies for cold-start item recommendation within the Netflix domain. We first assess the direct performance of the proposed approaches, and then systematically evaluate supervised fine-tuning, reinforcement learning-based methods, and their combined efficacy. Through extensive experiments on real-world data, we demonstrate that LLM reasoning delivers significantly superior performance in cold-start recommendation scenarios, where they can outperform Netflix’s production ranking model by 8\% in certain cases. 


In conclusion, our contributions can be summarized as follows:

\begin{itemize}[leftmargin=*]

    \item We propose novel reasoning strategies for cold-start item recommendation. While previous work has primarily focused on warm-start scenarios with sufficient user-item interactions, our approach leverages the inherent reasoning capabilities and extensive world knowledge of LLMs to better infer user preferences toward new or underrepresented items where classic collaborative filtering methods struggle due to limited interaction data.
    \item We conduct a comprehensive investigation of fine-tuning methodologies that goes beyond existing approaches by systematically examining both supervised fine-tuning and reinforcement learning fine-tuning strategies, as well as their combinations. This represents a significant advancement over current research that has largely focused on direct reasoning approaches or supervised methods alone, providing valuable insights into the comparative and combined effectiveness of different fine-tuning paradigms for LLM-powered recommendation systems.
    
\end{itemize}

\begin{figure*}[th]
\centering
\begin{minipage}{0.85\linewidth}
    \centering
    \includegraphics[width=0.9\linewidth]{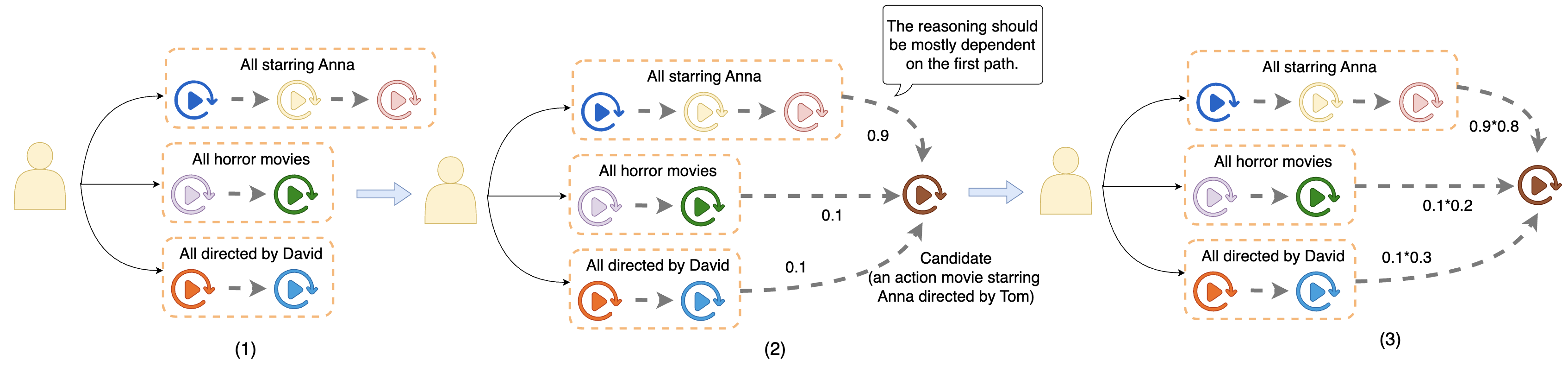}
\end{minipage}
\caption{Workflow of Structural Reasoning.}
\vspace{-1mm}
\label{fig:structural_reason}
\end{figure*}

\section{Methodology}
\subsection{Problem Definition}\label{sec:pd}
We focus on the scenario of cold-start item recommendation. Specifically, we take the cold-start items as those launched after the latest interaction of all user historical interactions used in our experiments.
Also, the launch of cold-start items is later than the training time of the base LLMs we used.

We evaluate all methods on the re-ranking task. Specifically, we selected the top 40 titles from 
the ranking model used in the production environment at Netflix.  We then further  sample 10 cold start titles including the target title, to form a total of 50 candidates. The final task is to re-rank these candidates for recommendation.

\subsection{Reasoning for Recommendation}

In this project, we proposed two reasoning strategies: Structural Reasoning and a variant of the Self-Consistency approach, targeting to infer users' preferences towards cold-start items with minimal prior knowledge. Both strategies are designed to decompose user interaction history into multiple reasoning paths, enabling a more accurate analysis of user preferences towards cold-start items.

\subsubsection{Structural Reasoning} 
Structural Reasoning instructs the LLM to systematically decompose structural information from a user's interaction history (e.g., watching history related to preference factors) into distinct reasoning paths. Then the model evaluates compatibility between each path and candidate items by computing match scores, which are then aggregated using weights based on factors like path prominence and recency. This enables the LLM to recommend the item with the highest overall score.
In summary, Structural Reasoning consists of three main steps:

\noindent \textbf{Reasoning Paths Construction.}  We first construct a graph of reasoning paths from the user's viewing history and provide it as an exemplar to guide the LLM in automatically generating similar structures during inference. Specifically, as illustrated in Figure \ref{fig:structural_reason}.1, we construct three reasoning paths based on the user's preferred actors, movie genres, and directors.

During inference, we prompt the LLM to identify the most significant factors influencing the user's current preferences, then construct a reasoning path for each factor by sequentially extracting relevant events from the user's interaction history.

\noindent \textbf{Factor Match Calculation.}
Second, for each candidate item, we instruct the LLM to compute a match score with respect to each constructed reasoning path. Specifically, we provide relevant candidate attributes, such as title, launch date, and genre, to guide the LLM in determining alignment and subsequently computing match scores against corresponding reasoning paths. As the case in Figure \ref{fig:structural_reason}, for a candidate that is an action movie starring Anna and directed by Tom, the reasoning path derived from the user's viewing history of movies starring Anna should receive the highest match score.

\noindent \textbf{Importance Weights and Aggregation.}
Finally, we instruct the LLM to assign importance weights to each reasoning path by explicitly considering factors such as factor prominence, action type, and recency. Using these importance weights, the LLM aggregates match scores for each candidate item, computing a weighted sum to generate an overall relevance score. Ultimately, the LLM ranks all candidate titles based on these aggregated scores, thereby producing the final recommendation list.


\subsubsection{Soft Self-Consistency}  The second strategy we proposed adapts the Self-Consistency approach~\cite{kumar2025llm}, which generates multiple parallel reasoning paths and aggregates them for final recommendations. In our implementation, the LLM autonomously constructs diverse reasoning paths based on key aspects of user preferences, deciding which aspects to prioritize. These paths are then synthesized into a final decision. However, rather than employing the classic self-consistency method's strict majority voting mechanism, we employ a soft summarization strategy, called Soft Self-Consistency, that allows the model to integrate and reconcile its reasoning paths before reaching a conclusion. This modification provides greater autonomy in leveraging reasoning capabilities and can yield performance gains over rigid aggregation policies that risk overlooking nuanced interactions between reasoning perspectives.

It is noteworthy that our Soft Self-Consistency approach can be viewed as the simplest variant of the proposed Structural Reasoning method. Specifically, Structural Reasoning explicitly instructs the LLM to construct reasoning paths by extracting historical interaction events guided by clearly identified preference factors. The model is then prompted to calculate factor match scores and assign importance weights to each reasoning path, ultimately determining the overall scores and rankings of candidate items. In contrast, the Soft Self-Consistency approach delegates both the construction of reasoning paths and the summarization of recommendations entirely to the LLM itself, without explicit instructions or regularization. Consequently, Structural Reasoning represents a more sophisticated and structured approach, while Soft Self-Consistency provides a simpler but more efficient method.

\subsection{LLM Fine-Tuning}

Beyond developing reasoning strategies, we further investigate the effectiveness of fine-tuning the base LLM. Specifically, we explored both Supervised Fine-Tuning and  Reinforcement Learning-based Fine-Tuning, as well as their combinations.

\subsubsection{Supervised Fine-Tuning}

Supervised fine-tuning (SFT) enhances LLM performance by learning from high-quality next-token prediction trajectories. In our studied context, we define reasoning paths that lead to successful user preference predictions as exemplary patterns for the model to emulate. However, since the ultimate objective is to improve the LLM's reasoning capabilities rather than merely replicate patterns from limited training data, we propose incorporating successful reasoning paths from multiple strategies to diversify the training corpus and mitigate overfitting risks. 


\subsubsection{Reinforcement Learning Fine-Tuning}

Reinforcement Learning Fine-Tuning (RLFT) has gained significant popularity due to its inherent flexibility in exploring beyond existing training data to achieve superior language generation performance. A critical component of RLFT is the reward function design, which provides supervisory signals to evaluate the quality of LLM-generated responses. In our recommendation context, the reward function should reflect the model's ability to generate recommendations that align with user preferences.
We define our reward function based on parsing the final recommendation output: a reward of 1 is assigned for correct predictions matching the ground truth, -0.1 for incorrect predictions, and -1 for parsing failures. Among the most popular RL methods that are widely adopted in both academia and industry,
we select GRPO \cite{shao2024deepseekmath} for its optimal balance between computational efficiency and performance. 


\subsubsection{SFT+RLFT}


In addition to evaluating SFT and RL fine-tuning separately, we also explored their combined application. We first fine-tuned the base LLM with SFT on demonstration reasoning trajectories, then further fine-tuned the saved model with GRPO using the defined reward function. After RL convergence, the resulting model integrated the benefits of both SFT and RLFT and was used for reasoning at inference.

\section{Experiment}

\subsection{Experimental Setup}

\subsubsection{Experiment Setting} 
The task setting and cold-start definition are illustrated in Section \ref{sec:pd}.
Specifically, the cold-start items never appear in user interactions within either the training or testing datasets and serve exclusively as ground-truth targets in the cold-start evaluation set. Crucially, their launch dates occur after the training cutoff of the base LLMs, ensuring they are genuinely new to the models. As a result, the LLMs must infer user preferences by reasoning over item characteristics using semantic information such as titles and genres, combined with their inherent knowledge.


For evaluation, we use Recall@1 for Discovery as the primary metric and Recall@1 for AnyPlay as the secondary metric. Discovery refers to plays involving novel or previously unseen content, while AnyPlay captures any type of user playback behavior.

\subsubsection{Data} To construct the cold-start test set, we sampled thousands of interactions from records within one day, ensuring that the target items (latest interactions) met the cold-start definition. For the SFT training set, we sampled 16,218 interactions after removing cold-start items. Applying our reasoning methods produced 7,252 successful reasoning paths—derived from both Base-Reason and Soft Self-Consistency, which achieved optimal performance on the AnyPlay and Discovery metrics, respectively.

For GRPO training, no demonstrated output trajectories are required, as the LLM generates these autonomously during training. Instead, we use prompts combining specific reasoning strategies with contextual information. To increase positive reward density and improve training stability, we oversampled prompts that led to successful recommendations by directly applying the corresponding reasoning strategies. This yielded 2,834 prompts for Soft Self-Consistency and 3,484 prompts for Base-Reason.

\subsection{Reasoning without Fine-Tuning}

\subsubsection{Baselines}

We evaluate our approaches against three baseline strategies. 
The first one, \textbf{Direct-Rec}, directly prompts the LLM to produce a final recommendation without any intermediate reasoning. \textbf{Base-Reason} guides the LLM through a two-step process: (i) summarizing the user’s interests and (ii) reasoning about preferences toward candidate items based on this summary to generate the final recommendation. The instruction prompt includes two few-shot examples, detailed guidelines, and illustrative reasoning patterns. \textbf{Fast-Reason} builds on Base-Reason by doubling the maximal interaction length, while simplifying the reasoning process through shorter example patterns and reducing context from few-shot to one-shot learning.

\begin{table}[ht]
\centering
\small
\caption{Relative performance (\%)  of  Fast-Reason (FR), {Direct-Rec (DR)}, {Base-Reason (BR)}, {Structural Reasoning (SR)}, and {Soft Self-Consistency (SSC)}.}


\begin{tabular}{p{0.95cm}
    >{\centering\arraybackslash}p{1cm} 
    >{\centering\arraybackslash}p{1cm} 
    >{\centering\arraybackslash}p{1cm}
    >{\centering\arraybackslash}p{1cm}
    >{\centering\arraybackslash}p{1cm}
    }
\hline
\textbf{Metric} & \textbf{FR} & \textbf{DR} & \textbf{BR} & \textbf{SR} & \textbf{SSC} \\
\hline
\small AnyPlay   & 0\% & +83.33\% & +104.17\% & +16.67\% & +4.38\% \\
\small Discovery & 0\% & -34.04\% & -29.15\% & -6.36\% & +6.38\% \\
\hline
\end{tabular}
\vspace{-2mm}
\label{tab:direct}
\end{table}

\subsubsection{Experimental Results}

We first conduct experiments comparing baselines to evaluate the effectiveness of different reasoning strategies for cold-start recommendation in our study. For all three approaches, we employ Qwen-2.5-32B-Instruct as the base LLM and assess performance on both AnyPlay and Discovery metrics. The results are demonstrate in Table~\ref{tab:direct}. 

\begin{table*}[th]
\centering
\small
\caption{Relative performance (\%) of fine-tuning approaches w.r.t. \textit{w/o SFT} baselines for Base-Reason and Soft Self-Consistency.}
\label{tab:main}
\begin{tabular}{lcccccccc}
\hline
\multirow{2}{*}{\textbf{Metric}} 
& \multicolumn{4}{c}{\textbf{Base-Reason}} 
& \multicolumn{4}{c}{\textbf{Soft Self-Consistency}} \\
\cmidrule(lr){2-5}\cmidrule(lr){6-9}
& \textbf{w/o SFT} & \textbf{+SFT} & \textbf{+GRPO} & \textbf{+SFT+GRPO} 
& \textbf{w/o SFT} & \textbf{+SFT} & \textbf{+GRPO} & \textbf{+SFT+GRPO} \\
\hline
AnyPlay   & 0\% & +8.16\% & +5.10\% & +7.14\% & 0\% & +22.45\% & +12.24\% & +8.16\% \\
Discovery & 0\% & +2.63\% & +18.42\% & +7.89\% & 0\% & 0.21\% & +6.00\% & 0.34\% \\
\hline
\end{tabular}
\end{table*}

Comparing Direct-Rec and Base-Reason, we find that Base-Reason outperforms Direct-Rec on both metrics, with improvements of 11.4\% on AnyPlay and 22.6\% on Discovery. This highlights the benefit of incorporating explicit reasoning processes rather than prompting LLMs to generate cold-start recommendations directly. Comparing Base-Reason and Fast-Reason further shows that extending historical context while simplifying demonstrative patterns yields a substantial 23.7\% gain on Discovery, though at the cost of reduced AnyPlay performance.

We then compare our two proposed strategies, Structural Reasoning and Soft Self-Consistency, against Fast-Reason, the strongest baseline on Discovery. The results reveal distinct performance trade-offs. Structural Reasoning, the more complex approach, improves AnyPlay by 16.7\% but decreases Discovery by 6.4\%. In contrast, Soft Self-Consistency, a simpler strategy, improves Discovery by 6.4\% while maintaining parity on AnyPlay. These findings suggest that complex reasoning strategies, which require LLMs to extract and analyze structural information, may enhance modeling of general user behaviors but risk imbalanced performance across metrics. Simpler approaches that grant greater flexibility in reasoning, however, appear to achieve more balanced outcomes.

In summary, among all strategies, Base-Reason achieves the strongest AnyPlay performance, while our Soft Self-Consistency approach delivers the best results on Discovery.

\subsection{Supervised Fine-Tuning}

We next evaluated the effectiveness of supervised fine-tuning (SFT). Specifically, we applied QLoRA with 4-bit quantization to fine-tune the base Qwen2.5-32B-Instruct model. As described earlier, we collected 7,252 reasoning paths that resulted in successful recommendations from both Base-Reason and Soft Self-Consistency to serve as training data. This design enhances the generalizability and diversity of the training corpus while mitigating bias toward specific reasoning patterns, since the objective of SFT is to strengthen general reasoning capabilities rather than replicate behaviors from limited data. Training was performed for a single epoch to prevent overfitting on the relatively small dataset. 

The results Table~\ref{tab:main} demonstrate that SFT significantly enhances AnyPlay performance without degrading Discovery metrics. Specifically, SFT achieves an 8.2\% AnyPlay improvement for Base-Reason and a substantial 22.4\% improvement for Soft Self-Consistency.

\subsection{RL-based Fine-Tuning}




Beyond supervised fine-tuning, we conducted experiments to explore the effectiveness of reinforcement learning-based fine-tuning. Specifically, while PPO was the first widely adopted RL method for LLMs with strong performance, it suffers from high computational cost due to its reliance on a large critic model. DPO-based methods simplify training by casting RL as supervised learning but rely on unrealistic assumptions about optimal solutions and have been shown to underperform in practice~\cite{xu2024dpo}. To address these limitations, we adopt GRPO, a recently proposed RL algorithm that preserves the standard RL framework while improving efficiency by estimating cumulative rewards through sampling, thereby avoiding both the need for a critic model and the shortcomings of DPO.

In addition to pure GRPO training, we also evaluated a hybrid strategy that combines SFT with RL: models were first fine-tuned using SFT, then further refined with GRPO.




From restults in Table~\ref{tab:main} on Base-Reason, we can see that GRPO improves both metrics, yielding 5.1\% on AnyPlay and 18.4\% on Discovery over the non-fine-tuned model. SFT alone delivers the largest AnyPlay gains, while SFT+GRPO achieves intermediate performance on both metrics. These results suggest that SFT primarily optimizes AnyPlay, whereas GRPO favors Discovery, and their combination struggles to reconcile these conflicting objectives, leading to balanced but not maximized performance.

A similar pattern emerges for Soft Self-Consistency. GRPO improves AnyPlay by 12.2\% and Discovery by 6.0\%, while SFT produces the highest AnyPlay gain (22.4\%). However, SFT+GRPO underperforms GRPO alone on both metrics, motivating further experiments in warm-start scenarios.

\begin{table}[ht]
\centering
\small
\caption{Warm-start performance of Soft Self-Consistency.}
\label{tab:grpo_3}
\begin{tabular}{p{0.95cm}
    >{\centering\arraybackslash}p{0.85cm} 
    >{\centering\arraybackslash}p{0.85cm} 
    >{\centering\arraybackslash}p{0.85cm}
    >{\centering\arraybackslash}p{1.45cm}
    >{\centering\arraybackslash}p{1.45cm}
    }
\hline
\textbf{Metric} & \textbf{w/o FT} & \textbf{+SFT} & \textbf{+GRPO} & \textbf{+SFT+GRPO} & \textbf{Production} \\
\hline
AnyPlay   & 0\% & +17.31\% & +11.54\% & +25.00\% & +96.15\% \\
Discovery & 0\% & +15.22\% & +17.39\% & +32.61\% & +23.91\% \\
\hline
\end{tabular}
\end{table}

As shown in Table~\ref{tab:grpo_3}, the SFT+GRPO approach achieves optimal performance in warm-start scenarios, outperforming SFT and GRPO individually on both AnyPlay and Discovery. Overall, our experiments indicate that SFT+GRPO performs strongly on warm-start tasks but struggles with cold-start recommendations. This outcome is intuitively reasonable given that the training data excludes cold-start items. In that case, the model that excels at recommending familiar items from the training corpus may struggle with previously unseen content.


Notably, in warm-start settings, 
\textbf{the improvement margin of SFT+GRPO over the baseline is 8\% higher than that of the Netflix’s production model on the Discovery metric}.
This result is striking given that the production model is trained on hundreds of millions of samples with sophisticated reward mechanisms refined through extensive empirical experimentation, whereas our approach relies on only thousands of samples and a simple reward design. Despite training with significantly fewer samples, SFT+GRPO achieves superior Discovery performance, highlighting a promising direction for future applications of LLM reasoning in Netflix’s recommendation systems.




\bibliographystyle{ACM-Reference-Format}
\bibliography{reference}

\end{document}